\documentclass[aps,prb,twocolumn,groupedaddress]{revtex4-1}%
\usepackage{amsmath,amssymb}
\usepackage{bm}
\usepackage{graphicx}
\usepackage{float}
\usepackage{amsmath, amssymb}
\usepackage{color}
\usepackage{amsmath}
\usepackage{amsfonts}
\usepackage{amssymb}%
\providecommand{\U}[1]{\protect\rule{.1in}{.1in}}
\begin{document}
\title{Gate-Controlled Semimetal-Topological Insulator Transition in an InAs/GaSb Heterostructure}
\author{Kyoichi Suzuki}
\email{suzuki.kyoichi@lab.ntt.co.jp}
\affiliation{NTT Basic Research Laboratories, NTT Corporation, 3-1 Morinosato-Wakamiya,
Atsugi, Kanagawa 243-0198, Japan}
\author{Yuichi Harada}
\affiliation{NTT Basic Research Laboratories, NTT Corporation, 3-1 Morinosato-Wakamiya,
Atsugi, Kanagawa 243-0198, Japan}
\author{Koji Onomitsu}
\affiliation{NTT Basic Research Laboratories, NTT Corporation, 3-1 Morinosato-Wakamiya,
Atsugi, Kanagawa 243-0198, Japan}
\author{Koji Muraki}
\affiliation{NTT Basic Research Laboratories, NTT Corporation, 3-1 Morinosato-Wakamiya,
Atsugi, Kanagawa 243-0198, Japan}

\date{\today}

\begin{abstract}
We report a gate-controlled transition of a semimetallic InAs/GaSb
heterostructure to a topological insulator. The transition is induced by
decreasing the degree of band inversion with front and back gate voltages.
Temperature dependence of the longitudinal resistance peak shows the energy
gap opening in the bulk region with increasing gate electric field. The
suppression of bulk conduction and the transition to a topological
insulator are confirmed by nonlocal resistance measurements using a dual
lock-in technique, which allows us to rigorously compare the voltage
distribution in the sample for different current paths without the influence
of time-dependent resistance fluctuations.

\end{abstract}


\maketitle

Topological insulators (TIs) have attracted strong interest as a new quantum
state of matter categorized neither as a metal nor an insulator.
\cite{Hasan,Qi,Ando,Kane1,Kane2,Bernevig-PRL,Bernevig-Science,Koenig-Science,Koenig-JPSJ,Roth}
The unique electronic properties of TIs originate from their band structures
characterized by conduction-valance band inversion with an energy gap opening
in the bulk region. The band inversion results from the interplay between the
alignment of bands with different orbital characters and the strong spin-orbit 
interaction inherent to materials containing heavy elements, such as Bi
compounds\cite{Hasan,Qi,Ando} and
HgTe.\cite{Bernevig-Science,Koenig-Science,Koenig-JPSJ,Roth} Owing to their
topological nature, gapless states with distinct transport characteristics
emerge at the surfaces and edges of three- and two-dimensional (2D) TIs,
respectively. In a 2D TI, also known as a quantum spin Hall insulator, the
edge state comprises counterpropagating channels with opposite spin.
\cite{Hasan,Qi,Kane1,Kane2,Bernevig-PRL,Bernevig-Science} Since elastic back
scattering is prohibited by time-reversal symmetry, dissipationless
spin-polarized transport is expected. The observation of the quantum spin Hall
effect in a HgTe/CdTe quantum well\cite{Koenig-Science,Koenig-JPSJ,Roth} has
established it as a prototypical system for 2D TIs.

Recently, InAs/GaSb heterostructures, with the so-called \textquotedblleft
broken-gap\textquotedblright\ or \textquotedblleft type-II\textquotedblright%
\ band alignment, have attracted interest as a route to achieving a 2D TI by
a combination of materials each having a trivial band structure typical of
III-V semiconductors.\cite{Liu,Knez-PRL1,Suzuki-PRB,Spanton,Knez-PRL2,Du} 
Different from
conventional TIs whose topological properties originate from the band
structure of individual materials, the band inversion in InAs/GaSb is provided
by the relative band alignment of the two semiconductors, where the InAs
conduction-band bottom is located below the GaSb valence-band top. Experiments 
have so far confirmed the existence of edge channels in InAs/GaSb heterostructures in
the band-inverted regime.\cite{Knez-PRL1,Suzuki-PRB,Knez-PRL2,Du,Spanton}
Additionally, it has been shown in Ref.~\onlinecite{Suzuki-PRB} that the band
alignment can be tuned by the thickness$\ w$ of the InAs layer. This is
manifested by the transport behavior's changing from that of a normal band
insulator through that of a TI to semimetallic as $w$ is varied from 10 to 14
nm. However, the major advantage of the InAs/GaSb system 
--the band alignment can be controlled in situ by gate voltages\cite{Liu,Naveh-APL}-- 
has
not yet been fully exploited.

In this paper, we study the effects of gate electric field on the transport properties of micrometer-sized dual-gate InAs/GaSb heterostructure device using nonlocal transport measurements. Different from HgTe/CdTe, in which both the electron
and hole wave functions are confined to HgTe and the band inversion is
determined by the HgTe layer thickness, the electron and hole
wave functions in InAs/GaSb are confined to different layers, which allows us to tune the
degree of band inversion in situ by applied gate voltages.\cite{Liu,Naveh-APL}
We demonstrate that an originally semimetallic heterostructure can be tuned
into a TI using gate voltages. Our results represent an important aspect of
the InAs/GaSb system, which is useful for studies of TIs.

Figure 1(a) shows schematically the band diagram of the system we study in
this paper. The InAs/GaSb heterostructure is sandwiched between Al$_{x}%
$Ga$_{1-x}$Sb barriers and has gate electrodes on both sides. We focus on the
band-inverted regime, that is, the situation where the thicknesses of the InAs
and GaSb layers are such that the bottom of the electron subband in InAs (with energy
$E_{\mathrm{e}0}$) is located below the top of the hole subband in GaSb (with
energy $E_{\mathrm{h}0}$) as shown in Fig.~1(c). The electron and hole wave functions, which are
predominantly confined to the InAs and GaSb layers, respectively, are
hybridized through the heterointerface to produce an anticrossing gap $\Delta$
at a finite in-plane wave vector $k$ [Fig.~1(b)]. As a consequence of the
inverted band alignment, linearly dispersing bands connecting the upper and
lower bands emerge [dotted lines in Fig.~1(b)], which correspond to the edge
channels. When the Fermi level is in the energy gap, the system becomes a 2D
TI, where the conductivity in the bulk region is expected to vanish and the
transport is governed by the edge channels. Experimentally, while the existence
of the anticrossing gap has been confirmed by capacitance
spectroscopy,\cite{Yang} transport measurements usually indicate a resistance
value that remains fairly low when the Fermi level is swept across the
expected gap region.\cite{Cooper,Pan,Nichele,Nguyen} One possible origin of this behavior is the
semimetallic band structure that results from the anisotropy of the GaSb
valence band\cite{Lakrimi} [Fig.~1(e)], which we will discuss later.

\begin{figure}[t]
\begin{center}
\includegraphics*[width=0.9\linewidth]{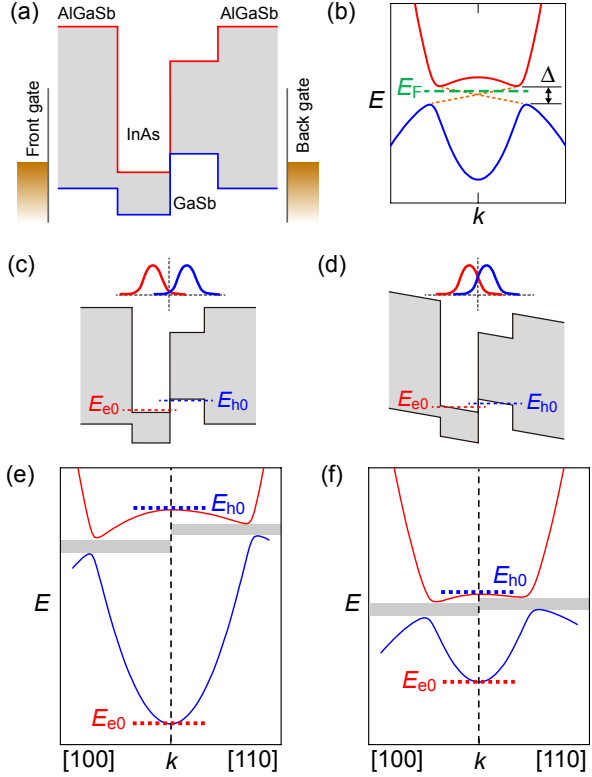}
\end{center}
\caption{
(Color online) 
Schematics of (a) band-edge profile of the sample used and (b) dispersion relations in a TI.
(c) and (d) Schematics illustrating the effects of gate electric field on the squared wavefunctions of electrons (red) and holes (blue) (upper panels) and band-edge profile (lower panels).
The diagrams in (c) and (d) represent the flat-band condition and the case with the gate electric field increasing the electron-hole wavefunction overlap, respectively.
$E_{\mathrm{e}0}$ and $E_{\mathrm{h}0}$ indicate the energies of the electron and hole ground subbands, respectively.
(e) and (f) Schematic illustrations of effects of gate electric field on the dispersion relation in the presence of valence-band anisotropy.
(e) and (f) correspond to the situation in (c) and (d), respectively. The shaded regions indicate the energy range of the anticrossing gap for each $k$ direction.
}
\label{f1}
\end{figure}

Figures~1(d) and (f) illustrate the effects of an electric field on the band
structure. Applying a positive voltage on the GaSb side decreases
$E_{\mathrm{h}0}$ while a negative voltage on the InAs side increases
$E_{\mathrm{e}0}$. This results in a lesser degree of band inversion as
measured by $E_{\mathrm{h}0}-E_{\mathrm{e}0}$. Consequently, the anticrossing
is shifted to a smaller $k$ and, at the same time, the larger spatial overlap
of the electron and hole wave functions leads to an increase in $\Delta
$.\cite{Naveh-APL} As we will show later, both of these effects help to
supress the bulk conductivity and stabilize the TI phase.

\begin{figure}[t]
\begin{center}
\includegraphics*[width=0.9\linewidth]{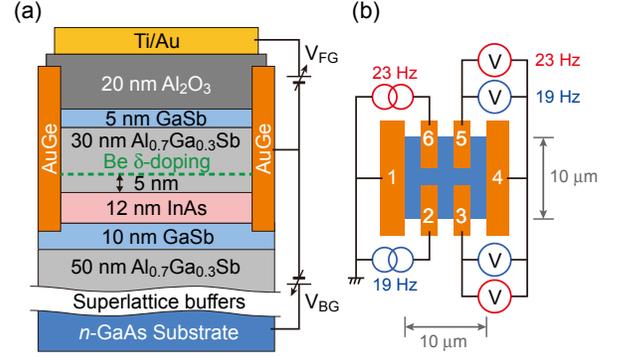}
\end{center}
\caption{
(Color online) (a) Schematic of sample structure.
(b) Six-terminal device pattern and typical nonlocal resistance measurement setup using dual lock-in technique.
The distance between adjacent contacts is $2$ $\mu$m.
The side contacts on the opposite edges (2--6 and 3--5) are separated by $2$~$\mu$m.
}
\label{f2}
\end{figure}

The InAs/GaSb heterostructure we used comprises InAs
and GaSb layers with nominal thicknesses of 12 and 10 nm, respectively, sandwiched between
Al$_{0.7}$Ga$_{0.3}$Sb barriers [Fig.~2(a)]. The structure was grown by
molecular beam epitaxy on an \textit{n}$^{+}$-GaAs substrate which serves as a
back gate. To bring the Fermi level close to the energy gap
region, the upper Al$_{0.7}$Ga$_{0.3}$Sb barrier is $\delta$-doped with
$[\mathrm{Be}]=5\times10^{11}$~cm$^{-2}$ at a setback of $5$~nm.\cite{Kadow} A small six-terminal
device with AuGe Ohmic contacts [Fig.~2(b)] is formed using deep
ultraviolet lithography and wet etching. The distance between adjacent
contacts is 2 $\mu$m. A Ti/Au front gate is evaporated on a 20-nm-thick
Al$_{2}$O$_{3}$ gate insulating layer deposited by atomic layer
deposition.\cite{Suzuki-APEX}

Transport measurements were performed using a lock-in technique at a
temperature of $T=0.27$ K, unless otherwise noted. For nonlocal resistance
measurements, alternating currents with different frequencies ($19$ and $23$
Hz) are injected from different contacts [Fig.~2(b)]. For each pair of voltage
probes, two lock-in amplifiers are used to monitor the two frequency
components. This dual lock-in setup allows us to simultaneously measure
nonlocal resistances for different current injection paths. As we will see
later, this method is particularly useful for the present purpose: it not only
saves measurement time, but also allows for rigorous comparison between
results for different current paths without its being influenced by irreproducible
time-dependent fluctuation of the sample characteristics.

Figure 3(a) shows longitudinal resistance $R_{14,23}$ at different
back-gate voltages ($V_{\text{BG}}$) plotted as a function of front-gate
voltage ($V_{\text{FG}}$). The suffixes of $R_{ij,kl}$ (= $V_{kl}/I_{ij}$)
indicate the contacts used for driving the current ($i,j$) and measuring the
resultant voltage ($k,l$). Because of the Be doping in our sample, the Fermi level
is in the valence band at $V_{\text{FG}}=0$ V. With increasing $V_{\text{FG}}%
$, the Fermi level moves toward the conduction band. The resistance curve
shows a peak when the Fermi level passes through the gap region. When the back
gate is unbiased ($V_{\text{BG}}=0$ V), the peak resistance is low ($<1$
k$\Omega$). As $V_{\text{BG}}$ is increased, the resistance peak shifts to
lower $V_{\text{FG}}$ and grows noticeably, reaching $\sim30$ k$\Omega$ at
$V_{\text{BG}}=10$ V. At the same time, spike-like fluctuations superimposed
on the peak become conspicuous. These fluctuations are mostly reproducible for
repeated gate sweeps, except for irreproducible slow time-dependent
ones.\cite{Suzuki-PRB}

\begin{figure}[t]
\begin{center}
\includegraphics*[width=0.9\linewidth]{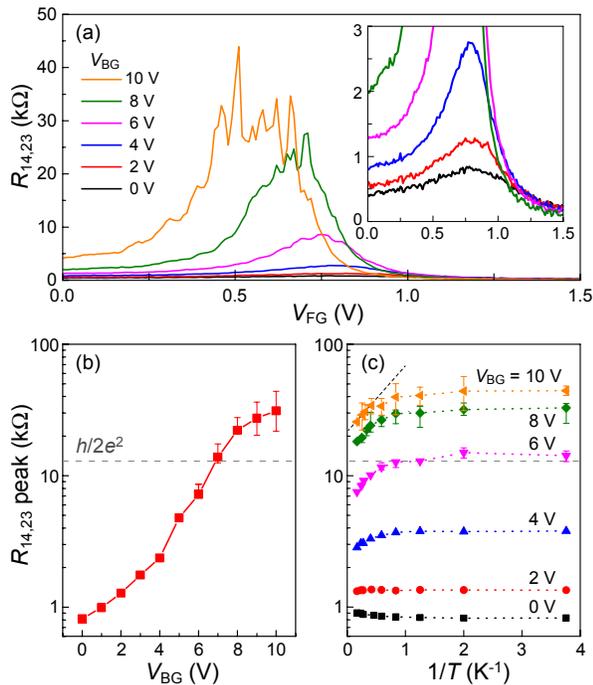}
\end{center}
\caption{
(Color online) (a) Longitudinal resistance ($R_{14,23}$) at different back gate voltages ($V_\text {BG}$) as a function of front gate voltage ($V_\text{FG}$).
Inset: magnified view of the the $R_{14,23}$ traces in the low-resistance regime.
(b) Peak value of $R_{14,23}$ obtained from Gaussian fitting of the $R_{14,23}$ traces in (a), plotted as a function of $V_\text{BG}$.
(c) Temperature dependence of $R_{14,23}$ peak at different $V_\text{BG}$. The data are plotted as a function of inverse temperature ($1/T$).
The short-dashed line is the Arrhenius fit to the data at $V_\mathrm{BG}=10$~V, with the slope corresponding to a gap of $0.2$~meV.
}
\label{f3}
\end{figure}

The height of the $R_{14,23}$ peak deduced by fitting it with a Gaussian is  plotted 
in Fig.~3(b) as a function of $V_{\text{BG}}$.\cite{Gaussian} The
horizontal dashed line in the figure indicates the value, $h/2e^{2}$ ($\sim13$
k$\Omega$), expected from the quantized edge transport for the geometry used
($h$: Planck's constant. $e$: elementary charge). At low $V_{\text{BG}}$, the
peak resistance is much lower than this value, indicating that considerable
bulk conduction exists even when the Fermi level is in the gap region. As
$V_{\text{BG}}$ is increased, the peak resistance increases significantly,
exceeding $h/2e^{2}$ at $V_{\text{BG}}\sim 7$ V, and then tends to saturate at
higher $V_{\text{BG}}$. Although edge transport in a 2D TI should be
protected from backscattering, in actual systems the conductance quantization
is disrupted by processes that equilibrate the counterpropagating edge
channels, as seen from the length dependence of the edge
conductance.\cite{Koenig-Science,Koenig-JPSJ,Du} The most probable cause is
the existence of electron and hole puddles coming from the spatial potential
fluctuations,\cite{Vayrynen-PRL,Vayrynen-PRB} which can terminate the edge
transport even in small samples and make the resistance higher than the
expected quantized value.\cite{Roth,Suzuki-PRB,Grabecki,Koenig-PRX}

Figure 3(c) shows the temperature dependence of the $R_{14,23}$ peak, plotted
as a function of inverse temperature. At $V_{\text{BG}}\leq2$ V, the peak
resistance is almost temperature independent, indicating that the system does
not have a full energy gap. At $V_{\text{BG}}= 4$~V the resistance is
thermally activated in the high-$T$ regime, suggesting the opening of an
energy gap. However, the temperature dependence saturates at low-$T$ regime
($\lesssim1$ K), and the system remains fairly conductive, with the resistance
well below $h/2e^{2}$. With increasing $V_{\text{BG}}$ further, the activated
behavior in the high-$T$ regime becomes more pronounced,\cite{Activation} and the resistance
value in the low-$T$ regime increases.

\begin{figure}[t]
\begin{center}
\includegraphics*[width=0.9\linewidth]{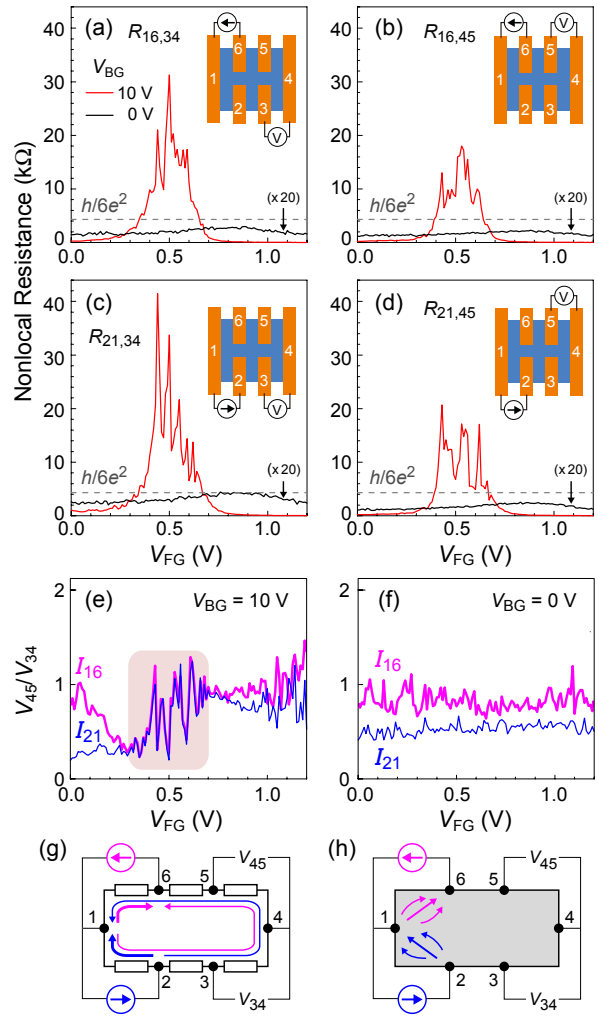}
\end{center}
\caption{
(Color online) (a)-(d) Nonlocal resistances (a) $R_{16,34}$, (b) $R_{16,45}$, (c) $R_{21,34}$, and (d) $R_{21,45}$, measured at $V_{\text{BG}}$ = 0 and 10 V as a function of $V_\text{FG}$.
(e) and (f) Ratios of nonlocal resistances (voltages) measured with adjacent voltage-probe pairs. Results for different current paths are compared for (e) $V_\text{BG}=10$ and (f) 0 V.
(g) and (h) Equivalent circuits for TI and normal conductor, respectively.
}
\label{f4}
\end{figure}

In our previous study, we compared samples with different
InAs layer thicknesses ($w=10$, $12$, and $14$ nm) while keeping
$V_{\text{BG}}=0$~V.\cite{Suzuki-PRB} The temperature dependence for $V_{\text{BG}}=2$~and $6$
V shown in Fig.~3(c) is similar, respectively, to that of the $w=12$ and $14$
nm samples (at $V_{\text{BG}}=0$ V) in the previous report, which
we identified as being a TI ($w=12$ nm) and a semimetal ($w=14$ nm). As seen
in Fig.~3(c), the resistance increases further at higher $V_{\text{BG}}$
($\geq8$~V). However, the behavior at $V_{\text{BG}}=8$ and $10$ V is distinct
from that of the normal band insulator previously reported for the $w=10$ nm
sample (at $V_{\text{BG}}=0$ V), in which the resistance increases to $\sim
$M$\Omega$. Here, it must be noted that the sample studied
here was fabricated from a wafer different from the one used for the $w=12$ nm
sample in the previous study. Comparing the results for the two
samples with $w=12$ nm, we infer that the actual InAs layer thickness is
larger in the sample used in this study.

The results in Fig.~3(c) show that the contribution of bulk conduction can be reduced by applied gate voltages.
The different contributions of bulk and edge transport at low and high
$V_{\text{BG}}$ can be further highlighted by nonlocal measurements. Figures~4(a)-(d)
display nonlocal resistances measured in four different configurations. The
black and red curves in each panel represent the results obtained for
$V_{\text{BG}}=0$ and $10$ V, respectively. The four traces for each
$V_{\text{BG}}$ were all measured simultaneously using the dual lock-in
technique. The horizontal dashed line in each panel indicate the value,
$h/6e^{2}$, expected for the quantized edge transport in the geometry used for
the nonlocal measurements.\cite{Suzuki-PRB} The nonlocal resistances for
$V_{\text{BG}}=0$ V (which are enlarged in the figures by a factor of $20$)
are far below this value, $h/6e^{2}$. Similar to the case of longitudinal
resistance, the nonlocal resistance peak grows significantly when
$V_{\text{BG}}=10$ V is applied, with the peak values exceeding $h/6e^{2}$.

As shown in Ref. \onlinecite{Suzuki-PRB}, that the transport is dominated by edge
channels can be confirmed by examining if the ratio of the nonlocal
resistances (or voltages) measured with adjacent voltage-probe pairs depends
on the current path. The results for $V_{\text{BG}}=10$ V are shown in
Fig.~4(e), where we compare the voltage ratios $V_{45}/V_{34}$ for two current
paths $I_{16}$ and $I_{21}$ as a function of $V_{\text{FG}}$. In the
$V_{\text{FG}}$ range corresponding to the peak in the nonlocal resistance
($0.3\lesssim V_{\text{FG}}\lesssim0.7$ V), the voltage ratios for the two
current paths completely agree with each other. Such agreement has been
confirmed for all contact geometries [Fig.~5(d) and Fig.~6 (Appendix)]. As shown in Ref.
\onlinecite{Suzuki-PRB}, these results indicate that the current flows only along
the edges of the sample, as illustrated in Fig.~4(g). This can be understood
by noting that if the current also flows through the bulk region, the voltage
distribution in the sample should depend on the current path [Fig.~4(h)]. In
fact, the latter behavior is seen in the low- and high-$V_{\text{FG}}$ regions
of Fig.~4(e), where the voltage ratios no longer agree with each other. This
shows that in these $V_{\text{FG}}$ ranges the central bulk region becomes
conductive, with the majority carrier type's being $p$ and $n$ on the low- and
high-$V_{\text{FG}}$ sides, respectively. We note that also in these $V_{\text{FG}}$ ranges the voltage ratio depends on $V_{\text{FG}}$, indicating
that edge transport coexists with bulk conduction. Similar results
were obtained from nonlocal resistance measurements at $V_{\text{BG}}=8$ V.

At $V_{\text{BG}}=0$ V [Fig.~4(f)], the voltage ratios are only a weak function of $V_{\text{FG}}$, 
and different current paths yield different voltage ratios.
Similar results were confirmed for other contact geometries. We note
that when the bulk transport is dominant the voltage distribution in the
sample is determined solely by the contact geometry and does not depend on
sample conductivity. Thus, the results in Fig.~4(f) indicate that at
$V_{\text{BG}}=0$ V the transport is dominated by bulk conduction even when
the Fermi level is in the expected gap region, implying a semimetallic band structure.

We study the evolution from a semimetal to a TI in more detail using nonlocal
measurements. Figures 5(a)-(d) display how the voltage ratios $V_{23}/V_{34}$
for two current paths $I_{16}$ and $I_{65}$ evolve with $V_{\text{BG}}$. With
increasing $V_{\text{BG}}\,$, the voltage ratios acquire $V_{\text{FG}}$
dependence, approaching each other, and  at $V_{\text{BG}}=10$ V they completely overlap 
in the range of $0.4\lesssim V_{\text{FG}}\lesssim0.7$ V. At
$V_{\text{BG}}=4$ V, the voltage ratios do not agree, indicating a finite
contribution of bulk conductivity. This is consistent with the longitudinal
resistance peak's being lower than $h/2e^{2}$ at $V_{\text{BG}}=4$ V [Fig.~3(c)]. The thermal activation of the longitudinal resistance in the
high-$T$ regime, in turn, indicates that at $V_{\text{BG}}=4$ V the bulk
conduction occurs locally. This implies that semimetallic and TI regions coexist
within the sample when the Fermi level is in the gap region. Spatial variation of the carrier density or layer thickness can cause such sample inhomogeneity.

\begin{figure}[t]
\begin{center}
\includegraphics*[width=0.9\linewidth]{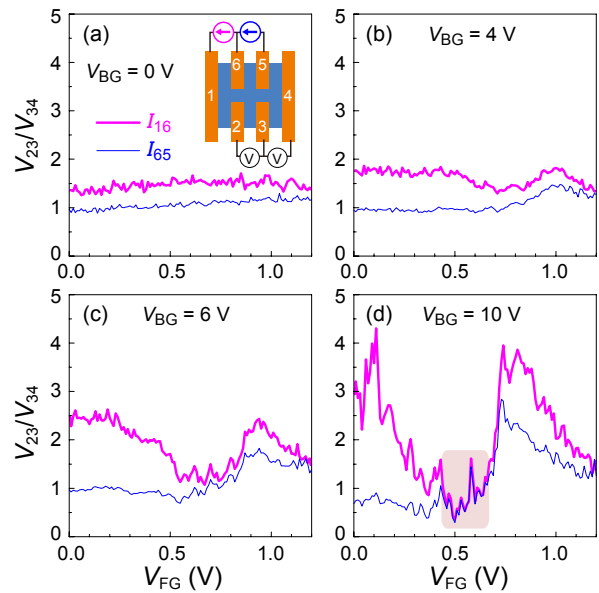}
\end{center}
\caption{
(Color online) Voltage ratios $V_{23}/V_{34}$ ($=R_{ij,23}/R_{ij,34}$) for current paths $I_{16}$ and $I_{65}$, measured at (a) $V_\mathrm{BG} = 0$, (b) 4, (c) 6, (d) 10 V.
}
\label{f5}
\end{figure}

Knez \textit{et al}.~also examined the effects of an electric field on 
residual bulk conductivity $g_{\mathrm{bulk}}$ in GaSb/InAs devices and
observed that $g_{\mathrm{bulk}}$\ decreased with applied gate
voltages.\cite{Knez-PRB,Knez-PRL1} However, in their undoped samples
$g_{\mathrm{bulk}}$\ remained significant ($\gtrsim2e^{2}/h$) even when the largest
electric field was applied.\cite{Knez-PRL1} The results were interpretted in
terms of the theory in Ref.~\onlinecite{Naveh-EPL}, which states that even weak
scattering can destroy the anticrossing gap and result in finite bulk
conductivity. For $\Gamma\ll\Delta\ll E_{\mathrm{h}0}-E_{\mathrm{e}0}$ (where $\Gamma$ is the level broadening), the theory provides a crude estimate of the residual conductivity: $g_{\mathrm{bulk}}\sim(e^{2}/h)(E_{\mathrm{h}0}-E_{\mathrm{e}0})/\Delta$.
The effect of the electric field is then understood to decrease $(E_{\mathrm{h}0}-E_{\mathrm{e}0})/\Delta$. However, the theory does not account for the vanishing of bulk conductivity we observed at high $V_{\text{BG}}$. This suggests that the assumption $\Gamma\ll\Delta\ll E_{\mathrm{h}0}-E_{\mathrm{e}0}$ no longer holds in the high-$V_{\text{BG}}$ regime of our experiments, where $E_{\mathrm{h}0}-E_{\mathrm{e}0}$ could be comparable to $\Delta$ depending on the InAs layer thickness.

Another possible mechanism that gives rise to residual bulk conductivity is the
anisotropy of the GaSb valence band.\cite{Lakrimi} In the presence of band anisotropy, the
energy position of the anticrossing gap depends on the $\mathbf{k}$ direction.
If the anisotropy is large as compared to $\Delta$, the anticrossing gaps for different $\mathbf{k}$ directions may not overlap in energy, resulting in a semimetallic band structure [Figs.~1(c) and (e)].
The effects of the electric field are to shift the
anticrossing to a smaller $k$, where the effect of the GaSb valence band
anisotropy is smaller [Fig.~1(d) and (f)], and enlarge $\Delta$%
.\cite{Naveh-APL} Both effects would work to suppress the bulk conductivity
and stabilize the TI phase. Most importantly, our nonlocal measurements
clearly show that the bulk conduction vanishes at higher $V_{\text{BG}}$ ($=8$
and $10$ V), demonstrating that an originally semimetallic heterostructure can be tuned
into a TI. Although further studies are necessary to clarify the role of
disorder,\cite{Du,Charpentier} our results clearly represent an important
aspect of the InAs/GaSb system, which will be useful for the study of TIs.

In summary, we have shown that a semimetallic InAs/GaSb heterostructure can be
tuned into a TI by applying gate voltages. Nonlocal resistance measurements
using a dual lock-in technique enabled us to confirm the suppression of bulk conductance
and the transition to a topological insulator without influence from
time-dependent resistance fluctuations. Our results warrant further
investigations to explore the transition between nontopological and topological insulators.

\begin{acknowledgments}
We sincerely thank Y. Ishikawa and H. Murofushi for
sample preparation. This work was supported by JSPS KAKENHI Grant Number 26287068.
\end{acknowledgments}


\appendix*
\section{}

\begin{figure}[b]
\includegraphics[width=1.0\linewidth,clip]{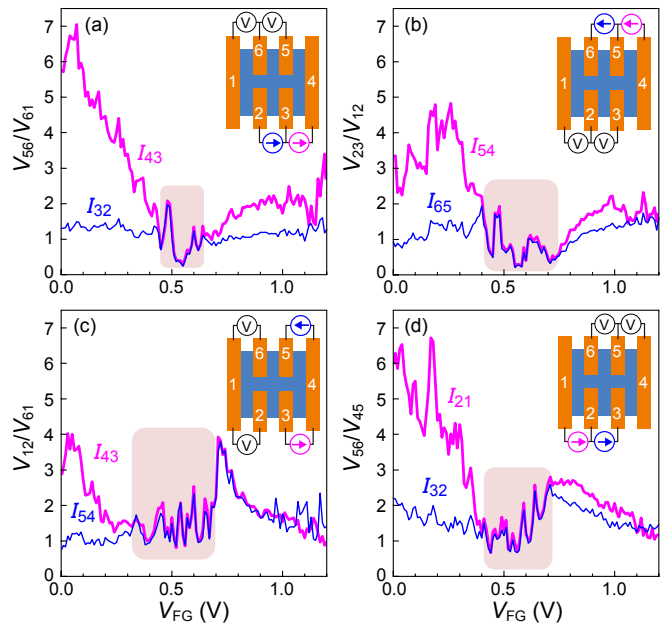}
\caption{
(Color online) 
Nonlocal resistance measurements at $V_\mathrm{BG} = 10$ V for various contact geometries shown in the insets. In each panel, nonlocal resistance ratios for different current paths are plotted as a function of $V_\mathrm{FG}$ ($T = 0.27$ K).
}
\label{fig6}
\end{figure}

Figure 6 summarizes the results of nonlocal resistance measurements at $V_\mathrm{BG} = 10$ V for contact geometries other than those in Figs.~4(e) and 5(d). 
For all the geometries, the nonlocal resistance ratios for different current paths precisely match each other in the shaded region, demonstrating that the bulk region is insulating and the transport is governed by edge channels in the corresponding $V_\mathrm{FG}$ range.


\begin{thebibliography}{99}
\bibitem{Gaussian}The Gaussian fitting is used simply to evaluate the mean height of the $R_{14,23}$ peak by effectively averaging the spike-like fluctuations that appear at high $V_\mathrm{BG}$. 
The error bars in Figs.~3(b) and 3(c) indicate the highest and lowest data points around the fitted peak.
\bibitem{Activation}Fitting the high-temperature section of the data at $V_\mathrm{BG}=10$~V with $R_{14,23}\propto \exp(\Delta/2k_\mathrm{B}T)$ provides the activation gap of $0.2$~meV, which is more than an order of magnitude smaller than the theoretical value.\cite{Liu} ($k_\mathrm{B}$ is the Boltzmann constant.) We consider that the gap reduction may be due to the band anisotropy discussed in this paper and/or disorder.
\end{thebibliography}
\end{document}